\documentstyle[prc,preprint,aps,graphicx,rotating,tighten]{revtex}

\begin{document}

\draft
\title{Novel Weak Decays in Doubly Strange Systems} 

\author{A. Parre\~no, A. Ramos}
\address{Departament d'Estructura i 
Constituents de la Mat\`eria, Universitat de Barcelona, Diagonal 647,
E-08028 Barcelona, Spain}

\author{C. Bennhold}

\address{Center for Nuclear Studies and Department of Physics, The George
Washington University, Washington DC, 20052, USA} 

\maketitle

\begin{abstract}
The strangeness-changing ($\Delta S = 1$) weak baryon-baryon
interaction is studied through the nonmesonic weak decay of
double-$\Lambda$ hypernuclei. Besides the usual nucleon-induced
decay $\Lambda N \to N N$ we discuss novel hyperon-induced
decay modes $\Lambda \Lambda \to \Lambda N$ and $\Lambda \Lambda
\to \Sigma N$.  These reactions provide unique access to the
exotic $\Lambda \Lambda$K and $\Lambda \Sigma$K vertices which
place new constraints on Chiral Pertubation Theory ($\chi$PT) 
in the weak
SU(3) sector.  Within a  meson-exchange framework, we use the
pseudoscalar $\pi,\eta,K$ octet for the long-range part while
parametrizing the short-range part through the vector mesons
$\rho, \omega, K^*$. Realistic baryon-baryon forces for the
$S=0,-1$ and $-2$ sectors account for the strong interaction in
the initial and final states. For $^6_{\Lambda \Lambda}$He 
the new hyperon-induced decay
modes account for up to 4\% of the total nonmesonic decay rate.
Predictions are made for all possible nonmesonic decay modes.
\end{abstract}

\section{Introduction}

The production and weak decay properties of strangeness-rich
systems is of fundamental importance for our understanding of
relativistic heavy-ion collisions and certain astrophysical
phenomena, such as neutron stars.  The simplest systems with
strangeness, hypernuclei with one or two bound $\Lambda$'s, have
been used to study both the strong and the weak baryon-baryon (BB)
interaction in the SU(3) sector.  Until now, hypernuclear weak
decay represents the only source of information on the $\mid \Delta
S\mid =1$ four fermion interaction where, in contrast to the weak
$\Delta S=0$ NN case, both the weak parity-conserving (PC) and
parity-violating (PV) amplitudes can be studied.

In the absence of exact solutions to low-energy QCD, effective field-theory
techniques based on chiral expansions have been fairly successful in the description
of hadronic observables in the (non-strange) SU(2) sector. The stability of the
chiral expansion is less clear for the SU(3) sector, due to the significant amount
of SU(3) symmetry breaking.  A well-known failure of SU(3) chiral perturbation
theory has been the prediction\cite{jenkins92} of the four PC
$P$-wave amplitudes in the weak nonleptonic decays of octet baryons, $Y \to N \pi$,
with $Y=\Lambda$, $\Sigma$ or $\Xi$.  Since large cancellations among tree-level
amplitudes are held responsible for the problem with the weak $P$-wave octet
amplitudes, it is imperative to assess whether this situation is universal within
SU(3) $\chi$PT or limited to a few exceptional cases.  However, other weak octet
baryon-baryon-meson (BBM) vertices can only be accessed through reactions that allow
for the virtual exchange of mesons, such as the reactions $\Lambda N \to N N$ and
$\Lambda \Lambda \to \Lambda(\Sigma) N$. The process $\Lambda N \to N N$ has been
extensively studied in an approach where the long-range part of this interaction is
based on the exchange of the SU(3) pseudoscalar meson octet ($\pi$, $K$, 
and $\eta$).
The pseudoscalar baryon-baryon-meson
vertices are considered fixed by experiment in the
case of the pion, and by SU(3) chiral algebra for the $K$ and the $\eta$.
Since the large momentum transfer in the reaction (typically 400 MeV/c) leads to a
mechanism where short-range effects must be included, they have been modelled either
through the exchange of the vector meson octet\cite{holstein,PRB97,PR01} ($\rho$,$\omega$
and $K^*$) or quark exchange\cite{SIO00}. The
vector baryon-baryon-meson vertices are constrained by much weaker SU(6)
considerations. The $\Lambda N \to N N$ process is then embedded in nuclear
many-body matrix elements using either correlated Faddeev amplitudes in the case of few-body
systems, hypernuclear shell
model wave functions or nuclear matter solutions within the Local
Density Approximation, depending on the mass number of the hypernucleus under
investigation.
While this description of hypernuclear weak decay is not as rigorous as
effective field theory would require, it nevertheless has been reasonably successful
in describing the available experimental data.

Since the late 1960's, the production and decay of single-$\Lambda$ hypernuclei has
been studied experimentally in great detail, but only very
few events involving doubly-strange objects have been reported
\cite{DJP66,MD63,SA91,YTI91,DMGD91}.
Double $\Lambda$ hypernuclei are produced via the $(K^-,K^+)$ reaction
at KEK (Japan) and BNL (USA), where a $\Lambda \Lambda$ hypernuclear fragment
can be formed by $\Xi^-$ capture on a nucleus.
The FINUDA experiment at
DA$\Phi$NE (Frascati, Italy) can produce double-$\Lambda$ hypernuclei by
stopping slow $K^-$ (coming from the $\Phi$ decay) into thin targets
to obtain data with higher energy resolution.
Studying the weak decay of those objects opens the door to a number of new exotic
$\Lambda$-induced decay modes: $\Lambda \Lambda \to \Lambda n$ and $\Lambda
\Lambda \to \Sigma N$. Both of these decays would involve hyperons in
the final state and should be distinguishable from the ordinary $\Lambda N \to
 NN$ mode.  The $\Lambda \Lambda \to \Lambda n$ channel is especially
intriguing since the dominant pion exchange is forbidden, thus forcing this
reaction to occur mostly through kaon exchange. One would
therefore gain access to the weak $\Lambda \Lambda$K and $\Lambda \Sigma$K
vertices.

In this paper, we extend previous weak decay calculations of single-$\Lambda$
hypernuclei into the $S=-2$ sector, thus exploring the power
of the $\Lambda \Lambda \to \Lambda (\Sigma) N$ process to shed light on the
novel weak vertices.
In order to take into account the effects of the strong interaction between the baryons,
correlated wave functions are
obtained from a $G$-matrix calculation for the initial $\Lambda$N and $\Lambda \Lambda$ 
states, while
a $T$-matrix equation is solved for the final NN and YN states using the Nijmegen
interaction models\cite{nij99}, in particular the NSC97f one.
       
\section{Decay rate}

In the weak nonmesonic decay of double-$\Lambda$ hypernuclei,
new hyperon-induced mechanisms, the $\Lambda \Lambda \to \Lambda N$ and the
$\Lambda \Lambda \to \Sigma N$ transitions (denoted as $\Lambda \Lambda
\to YN$ throughout the text) become possible in addition
to the dominant $\Lambda N \to NN$ decay mode.
Assuming the initial hypernucleus to be at rest,
the NMD rate is given by:

\begin{eqnarray}
\Gamma_{\rm nm} &=& \int  \frac{d^3 k_1}{(2\pi)^3}
 \int  \frac{d^3 k_2}{(2\pi)^3} \nonumber \\
&\times& \sum_{^{M_I  \{R\} }_{\{1\} \{2\}}}
\,\, (2\pi) \,\, \delta(M_H-E_R-E_1-E_2)
\,\, \frac{1}{(2J+1)} \,\,
\mid {\cal M}_{i \to f} \mid^2  \ ,
\label{eq:rate0}
\end{eqnarray}
where the quantities
$M_H$, $E_R$, $E_1$ and $E_2$ are the mass of the hypernucleus,
the energy of the residual \mbox{$(A-2)$}-particle system, and the total
asymptotic energies of the emitted baryons, respectively.
The integration variables ${\vec k}_1$ and ${\vec k}_2$ stand for
the momenta of the two baryons in the final state.
In the expression above,
the momentum-conserving delta function has been used to
integrate over the momentum of the residual nucleus.
The sum, together with the factor $1/{(2J+1)}$, indicates an average
over the initial hypernucleus spin
projections, $M_I$, and a sum over all quantum numbers
of the residual
\mbox{$(A-2)$}-particle system, $\{R\}$, as well as the spin and isospin
projections of the emitted particles, $\{1\}$ and $\{2\}$.
The total nonmesonic decay rate
can be written as:
\begin{equation}
\Gamma_{\rm nm}=\Gamma_{\rm NN} + \Gamma_{\rm YN} \ ,
\end{equation}
where the rate corresponding to $\Lambda N \to NN$
transitions
\begin{equation}
\Gamma_{\rm NN}=\Gamma_{\rm n}+\Gamma_{\rm p} \, \ ,
\end{equation}
is divided into 
a neutron-induced rate,
$\Gamma_{\rm n}:$ \mbox{$\Lambda n \to nn$} and a proton-induced one,
$\Gamma_{\rm p}:$ \mbox{$\Lambda p \to np$},
while the $\Lambda$-induced transitions
$\Lambda \Lambda \to YN$ give
rise to $\Lambda$n, $\Sigma^0$n and $\Sigma^-$p final states, hence:
\begin{equation}
\Gamma_{\rm YN} = \Gamma_{\Lambda {\rm n}} +
 \Gamma_{\Sigma^0 {\rm n}} + \Gamma_{\Sigma^- {\rm p}} \ .
\end{equation}

Note that while the individual rates have been written above as exclusive
observables, in principle, charge-exchange final state interactions (FSI) with the
residual nucleus obscurs a clean experimental discrimination of these channels on
the basis of the charge of the emitted particles. Monte Carlo intranuclear cascade
models are necessary \cite{ramos97} to extract the partial weak decay rates of
hypernuclei from experiment \cite{outa00}. The impossibility of measuring the
partial rates directly from the charge of the emitted particles was also
shown in the case of the weak decay of the hypertriton
\cite{hyptriton97}, a system where the
strong interaction effects can be treated
exactly using Fadeev equations with a realistic baryon-baryon potential.

In order to draw conclusions
regarding the weak dynamics, one has to write the hypernuclear 
transition amplitude, ${\cal M}_{i \to f}$,
in terms of the elementary two-body transitions, $B_1 B_2 \to B_3 B_4$,
and nuclear structure details have to be
treated with as few approximations and ambiguities as possible.
We work in a shell-model framework, hence the $\Lambda$ hyperons
and the nucleons are described by single-particle orbitals. In
addition, we assume a weak coupling scheme by virtue of which the
$\Lambda$ hyperons couple only to the ground state of the nuclear
core. Therefore, in the case of 
the $^6_{\Lambda \Lambda}$He hypernucleus studied here, with
quantum numbers \mbox{$J_I=M_I=0$},
\mbox{$T_I=M_{T_I}=0$}, the state will be given by
\begin{equation}
|^6_{\Lambda \Lambda}{\rm He} \rangle ^{J_I=M_I=0}_{T_I=M_{T_I}=0} =
|\Lambda \Lambda
\rangle^{J_\Lambda=M_\Lambda=0}_{T_\Lambda=M_{T_\Lambda}=0} \otimes
| ^4{\rm He} \rangle^{J_c=M_c=0}_{T_c=M_{T_c}=0}  \ ,
\label{eq:decomposition}
\end{equation}
where antisymmetry forces the two $\Lambda$ hyperons to be in a
$^1S_0$ state, since they are assumed to be in the lowest s-shell before
the decay occurs. This is so because, in general, 
hypernuclei with $\Lambda$'s in excited orbitals will
rapidly decay into the ground state through electromagnetic or
strong de-excitation processes, which are orders of magnitude
faster than those mediated by the weak interaction. 
The single-particle orbitals for nucleons and $\Lambda$'s
are taken as harmonic oscillator states with parameters $b_N=1.4$ fm and
$b_\Lambda=1.6$ fm, respectively. 
The nucleon parameter is chosen to account for the $^4$He form factor. 
The parameter for the $\Lambda$ wave function reproduces 
the Hartree-Fock (HF) probability of finding the two $\Lambda$ particles at 
relative distance $r$, obtained in Ref.~\cite{caro} by adjusting their model
parameters to the binding energies of the three observed double 
$\Lambda$-hypernuclei, $^6_{\Lambda \Lambda}$He, $^{10}_{\Lambda \Lambda}$B
and $^{13}_{\Lambda \Lambda}$B \cite{DJP66,MD63,SA91,YTI91,DMGD91}. 

To obtain the rate corresponding to the $\Lambda N \to NN$ transition, with
the initial $N$ being $n (t_N=-1/2)$ or $p (t_N=1/2)$, we
have to write the non-strange nuclear core as one nucleon coupled to a 3-particle 
system (with 
quantum numbers $J_3, M_3, T_3$ and $M_{T_3}$)
and decouple one of the
two $\Lambda$'s in the cluster. Therefore, the initial $\Lambda$N
pair will convert into a final NN pair with quantum numbers ${\vec
k}_1 s_1 t_1,~{\vec k}_2 s_2 t_2$, leaving a residual 4-particle system
with quantum numbers $J_R,M_R,T_R,M_{T_R}$. Performing all the necessary
decoupling and recoupling operations, we finally arrive at:
\begin{eqnarray}
{\cal M}_{\Lambda N \to NN} &=& 
\langle {\vec k}_1 s_1 t_1, {\vec k}_2 s_2 t_2; ^4_\Lambda 
{\rm A}^{J_R M_R}_{T_R M_{T_R}}| \, {\hat O} \, | \,
^6_{\Lambda \Lambda}{\rm He} \rangle \, | \frac{1}{2} \, -\frac{1}{2}
\rangle_\Lambda 
\nonumber \\
&=& \sum_{S} \sum_{T} \sum_{m_N} \sum_{S_0} \, 
\langle T M_T | \frac{1}{2} t_1, \frac{1}{2} t_2 \rangle \, 
\langle T M_T | \frac{1}{2} \, -\frac{1}{2}, \frac{1}{2} t_N
\rangle
\nonumber \\
&\times& 
\langle T_3  M_{T_3}, \frac{1}{2} t_N | T_c=0 \, M_{T_c}=0 \rangle  \,
\langle J_3 M_3,  \frac{1}{2} m_N | J_c=0 \, M_c=0 \rangle
\nonumber \\
&\times&
\langle \frac{1}{2} \, (m_N-M_{S_0}), J_3 M_3 | J_R M_R \rangle \,
\langle S  M_{S} | \frac{1}{2} s_1, \frac{1}{2} s_2 \rangle
\nonumber \\
&\times& 
\langle \frac{1}{2} \, (M_{S_0}-m_N),  \frac{1}{2} \, (m_N -  M_{S_0})|
J_\Lambda=0 \,
M_\Lambda=0 \rangle \nonumber \\ 
&\times&  
\langle \frac{1}{2} \, (M_{S_0}-m_N),  \frac{1}{2} m_N | S_0 M_{S_0} \rangle
\nonumber \\
&\times& 
\langle \vec{K} | \Psi^{\rm CM}_{\Lambda {\rm N}}\rangle \,
\langle \vec{k}, S M_S, T M_T| \,{\hat O} \,| \, \Psi^{\rm 
rel}_{\Lambda {\rm N}},
S_0 M_{S_0}, T M_T \rangle \ , 
\end{eqnarray}
where $T_3=1/2$, $M_{T_3}=-t_N$, $J_3=1/2$, $M_3=-m_N$, 
$M_{T_R}=-t_N$, $M_{S_0}=-M_R$, $M_S=s_1+s_2$ and 
$M_T=t_N-1/2=t_1 + t_2$. We
have assumed the $\Lambda$ at the weak vertex to behave as a
$|\frac{1}{2} -\frac{1}{2} \rangle$ isospurion in order to
naturally incorporate the experimentally observed $\Delta I = 1/2$ rule.
The momentum
states $\vec{k}_1$ and $\vec{k}_2$ have been transformed to total
($\vec{K}=\vec{k}_1 +
\vec{k}_2$) and relative ($\vec{k}=(\vec{k}_1 - \vec{k}_2)/2$) momenta and
the final two-nucleon state, $\langle \vec{k}, S M_S, T M_T |$,
must be properly antisymmetrized, which means that
the factor $(1-(-1)^{L+S+T})/\sqrt{2}$ will appear when a
decomposition in partial waves is performed for the outgoing plane wave.

For the $\Lambda \Lambda \to YN$ transition we don't need to 
decouple one of the 
hyperons from the cluster, neither a nucleon from the core. The residual 
4-particle system, which coincides with the $^4_\Lambda$He core in Eq. 
(\ref{eq:decomposition}), contains no strangeness, while the final two-particle
state contains one hyperon 
that can be either a $\Lambda$ ($|Y t_Y\rangle = |0 0\rangle$), a $\Sigma^-$
($|Y t_Y\rangle = |1 -1\rangle$) or a
$\Sigma^0$ ($|Y t_Y\rangle =| 1 0\rangle$). The $\Lambda \Lambda \to YN$ hypernuclear
amplitude
is given by:
\begin{eqnarray}
{\cal M}_{\Lambda\Lambda \to Y N} &=&
\langle{\vec k}_{N} s_{N} t_{N}, {\vec k}_Y s_Y t_Y; \, ^4{\rm
He} \,
| \, {\hat O} \, | \, ^6_{\Lambda \Lambda}{\rm He} \rangle \, |\frac{1}{2}
\, -\frac{1}{2} \rangle_\Lambda \nonumber \\
&=& \sum_{S,M_S} \langle \frac{1}{2} s_N, \frac{1}{2} s_Y | S M_S \rangle
\, 
\langle \frac{1}{2} t_{N}, Y t_Y | T M_T \rangle 
\nonumber \\
&\times &
\langle {\vec K} | \Psi^{\rm CM}_{\Lambda\Lambda}\rangle \, \langle {\vec k}, S
M_S, T M_T| \, {\hat O} \, | \Psi^{\rm rel}_{\Lambda\Lambda},
S_0 M_{S_0}, T_0  M_{T_0} \rangle  \ , 
\end{eqnarray}
where $S_0=M_{S_0}=0$, $T=T_0=1/2$ and $M_T=M_{T_0}=-1/2$.

\section{The Meson Exchange Potential}

\begin{figure}
\begin{center}
\includegraphics[width=10cm]{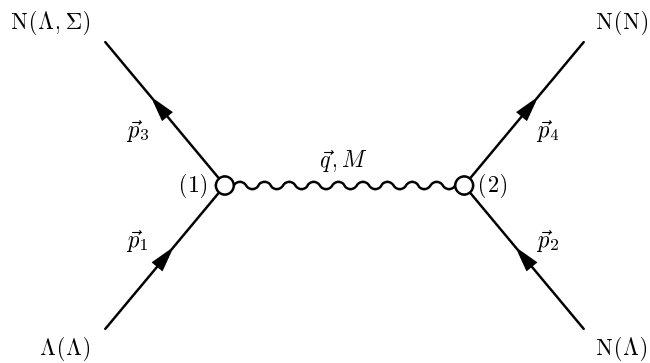}
\caption{Feynman diagram for the weak $\Lambda N \to NN$ and $\Lambda \Lambda
\to YN$ ($\Sigma$N or $\Lambda$n interaction).}
\label{fig:feydia}
\end{center}
\end{figure}

In a meson-exchange description, the $\Lambda N \to N N$ and $\Lambda \Lambda \to YN$
transitions are assumed to proceed via
the exchange of virtual mesons belonging to the ground-state
pseudoscalar and vector meson octets. The amplitude, which is schematically
represented by the Feynman diagram depicted in Fig.~[\ref{fig:feydia}],
reads
\begin{equation}
{\cal M}_{\rm M} = \int d^4 x \,\, d^4 y \,\, {\overline \Psi}_{p_3} (x)
\,\, \Gamma_1 \,\, \Psi_{p_1} (x) \,\, \Delta_{\rm M} (x-y)
\,\, {\overline \Psi}_{p_4} (y) \,\, \Gamma_2 \,\, \Psi_{p_2} (y) \ ,
\label{eq:uncF0}
\end{equation}
where $\Psi_{p} (x) = {\rm e}^{-{\rm i} p x} u (p)$
is the free baryon field of positive energy, $\Gamma_i$ the Dirac
operator characteristic of the baryon-baryon-meson (BBM) vertex and
$\Delta_{\rm M} (x-y)$ the meson propagator.

In Table \ref{tab:vert} we show the strong and weak hamiltonians
for pseudoscalar (PS) and vector (V)
mesons. The constants $A$, $B$, $\alpha$, $\beta$ and $\epsilon$ correspond
to the weak couplings, while $g$ ($g^{\rm V},g^{\rm T}$) represents the strong
(vector, tensor) one.

The nonrelativistic reduction of this amplitude gives us
the potential in momentum space \cite{PRB97}, which for
pseudoscalar mesons reads:

\begin{equation}
V_{ps}({\vec q}\,) = - G_F m_\pi^2
\frac{g}{2 M_2} \left(
{\hat A} + \frac{{\hat B}}{2 M_1}
{\vec \sigma}_1 \, {\vec q} \,\right)
\frac{{\vec \sigma}_2 \, {\vec q}\, }{{\vec q}^{\; 2}+\mu^2} \ ,
\label{eq:pion}
\end{equation}
where $G_F {m_\pi}^2=2.21 \times 10^{-7}$ is the Fermi weak constant,
${\vec q}$ is the momentum carried by the meson directed towards the
strong vertex, $\mu$ the meson mass and
$M_2$ ($M_1$) is the average of the baryon masses at the
strong (weak) vertex.
For vector mesons the potential reads:
\begin{eqnarray}
{V_{v}}({\vec q}\,)  &=&
G_F m_\pi^2
 \left( g^{\rm V}{\hat \alpha} - \frac{({\hat \alpha} + {\hat \beta} )
 ( g^{\rm V} + g^{\rm T})} {4 M_1 M_2}
({\vec \sigma}_1 \, \times {\vec q} \,)
({\vec \sigma}_2 \, \times {\vec q} \,) \right. \nonumber \\
& & \phantom { G_F m_\pi^2 A }
\left. - {\rm i} \frac{{\hat \varepsilon} ( g^{\rm V} + g^{\rm T})} {2 M_2}
({\vec \sigma}_1 \, \times
{\vec \sigma}_2 \, ) {\vec q} \,\right)
\frac{1}{{\vec q}^{\; 2} + \mu^2} \ .
\label{eq:rhopot}
\end{eqnarray}
In Eqs.~(\ref{eq:pion}), (\ref{eq:rhopot}) the operators
${\hat A}$, ${\hat B}$, ${\hat \alpha}$, ${\hat \beta}$ and 
${\hat \varepsilon}$ contain, apart from the weak coupling constants, 
 the specific isospin dependence of the potential, 
which is 
${\vec \tau}_1 \, {\vec \tau}_2$ for the isovector $\pi$ and $\rho$ mesons,
${\hat 1}$ for the isoscalar 
$\eta$ and $\omega$ mesons and 
a combination of both operators for the isodoublet 
$K$ and $K^*$.

Ref.~\cite{PRB97} introduces a compact way to write the transition
potential
in ${\vec r}$-space:
\begin{equation}
V({\vec r}\,) = \sum_{i} \sum_\alpha V_\alpha^{(i)}
({\vec r}\,) = \sum_i \sum_{\alpha}
V_\alpha^{(i)} (r) \, \hat{O}_\alpha ({\hat r}) \, \hat{I}_\alpha^{(i)} \, \ ,
\label{eq:compact}
\end{equation}
where the index $i$ runs over the different mesons exchanged ($i=1,\dots,
6$ represents $\pi,\eta$,K,$\rho,\omega$,K$^*$) and
$\alpha$ over the different
spin operators: central spin independent (C), central spin dependent (SS), tensor (T)
and parity violating (PV)
\[
{\hat O}_\alpha ({\hat r}) \, = \left\{
\begin{array}{ll}
\hat{1} & \mbox{C (only for vector mesons)} \, {\rm ,} \\
{\vec \sigma}_1 {\vec \sigma}_2  & \mbox{SS} \, {\rm ,} \\
S_{12} ({\hat r}) = 3 {\vec \sigma}_1 {\hat r}{\vec \sigma}_2 {\hat r} - 
{\vec \sigma}_1 {\vec \sigma_2} & \mbox{T} \, {\rm ,} \\
{\rm i} \, {\vec \sigma}_2 {\hat r} & \mbox{PV (pseudoscalar mesons)}\, {\rm ,}  \\
\left[ {\vec \sigma}_1 \times {\vec \sigma}_2 \right] {\hat r}& \mbox{PV 
(vector mesons)}  \, {\rm .}\\
\end{array}
\right. 
\]

In order to account for the finite size and structure of the particles 
involved in the process, form factors (FF) are included.
In our previous calculations of the decay of single-$\Lambda$ hypernuclei
\cite{PRB97} we used a monopole FF,
$F({\vec q}^{\,2})= ({\Lambda_i}^2-\mu^2)/({\Lambda_i}^2+{\vec q} ^{\, 2})$,
at both the weak and strong vertices,
with different cut-offs depending on the exchanged meson, $\Lambda_i$. 
Those cut-offs were taken from the J\"ulich YN interaction
model \cite{juelich}.
The latest version of the Nijmegen potentials\cite{nij99}, which is 
used here, also gives 
meson-dependent cut-offs  but uses an exponential FF at each vertex of the
type $F({\vec q}^{\,2})= {\rm exp}(-{\vec q}^{\, 2}/2 {\Lambda_i}^2)$. 
Therefore, the results presented here are also
obtained with the exponential-type FF, although, for technical reasons,
it is matched to a function of the type
$\widetilde{\Lambda}_i^2/(\widetilde{\Lambda}_i^2 +
\vec{q}\,^2)$ at $\mid \vec{q}\, \mid \simeq 400$ MeV/c, the most relevant
momentum transfer in the weak decay transition. By
construction, this type of monopole FF is equivalent to the exponential
one at $\vec{q}=0$ and we have verified that, 
up to a momentum transfer of
600 MeV/c above which
the transition amplitude is negligible,
the differences between both functions are less than 2\%.
The modified {\it cut-off}s, $\widetilde{\Lambda}_i$, to be used in the
monopole-type FF 
are listed in Table \ref{tab:ff}.

\begin{figure}[hbt]
\begin{center}
\includegraphics[width=12cm]{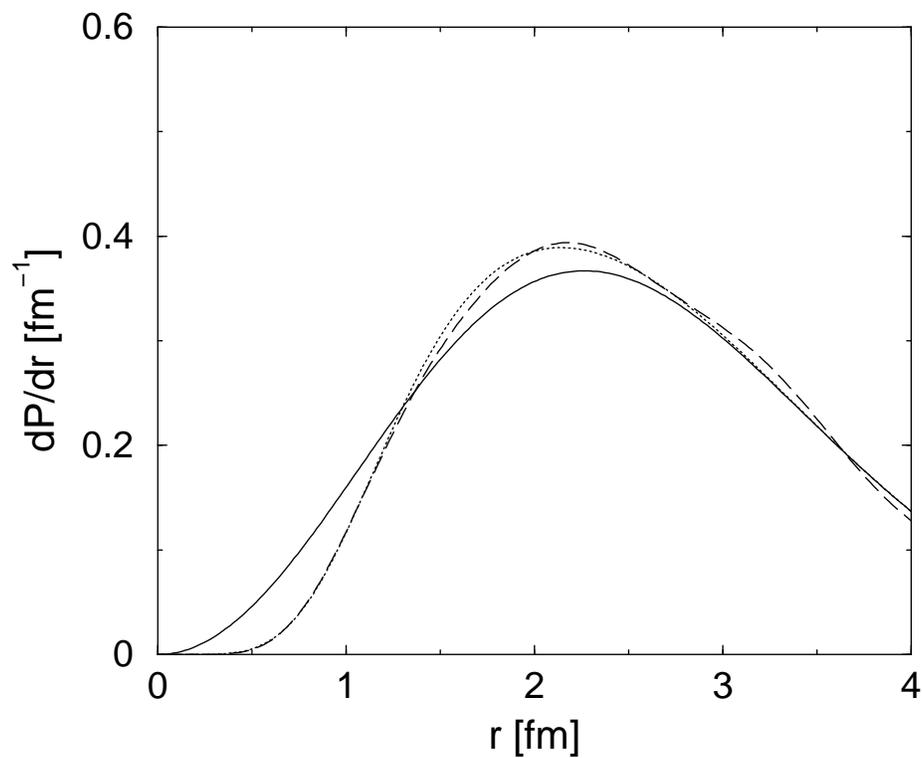}
\caption{Probability per unit length of finding two $\Lambda$'s at
a relative distance $r$.}
\label{fig:prob}
\end{center}
\end{figure}

To incorporate the effects of the strong baryon-baryon (BB) interaction 
we solve a T-matrix 
scattering equation in momentum space for the outgoing two-particle system 
(NN or YN). For the initial two-particle system ($\Lambda$N or
$\Lambda \Lambda$) we take into account medium effects, thus
 intermediate states only propagate into states allowed by
the Pauli operator ($G$-matrix).
For the initial interacting $\Lambda$N pair we previously found \cite{sitges}
that multiplying the independent two-particle
wave function by a spin-independent correlation function
of the type: 
\begin{equation}
f(r)=\left( 1 - {\rm e}^{-r^2 / a^2} \right)^n + b r^2 {\rm
e}^{-r^2 / c^2} \ ,
\label{eq:fsief}
\end{equation}
with $a=0.5$ fm, $b=0.25$ fm$^{-2}$, $c= 1.28$ fm, $n= 2$,
produced a correlated wave function which averaged the ones
obtained from a microscopic finite-nucleus $G$-matrix
calculation\cite{halder} using the soft-core and hard-core Nijmegen models
\cite{nij89}. 
In the case of the interacting $\Lambda\Lambda$ system, we again assume a
correlation function of the type of Eq.~(\ref{eq:fsief}). The parameters,
$a=0.80$ fm, $b=0.12$ fm$^{-2}$, $c=1.28$ fm and $n=1$, are determined to
reproduce the ratio between the correlated and uncorrelated $^1S_0$ $\Lambda\Lambda$
wave functions in nuclear matter at saturation density,
$\Psi_{^1S_0}(k,r)/j_0(kr)$, 
taking $k=100$ MeV/c as a representative momentum of the two
$\Lambda$'s in a finite nucleus. 
The nuclear-matter correlated wave function, $\Psi_{^1S_0}(k,r)$, has been 
obtained from a coupled-channel G-matrix calculation
using the NSC97f model of the new Nijmegen potentials.
In Fig.~\ref{fig:prob}, we show the probability per unit length of finding
two $\Lambda$ particles at a distance $r$, given by $ 4 \pi r^2
|\Psi^{\rm rel}_{\Lambda\Lambda}|^2 $. The solid line corresponds to 
the uncorrelated two-$\Lambda$ wave function, which in this work is
given by the 1s harmonic oscillator wave function with the parameter
$b_{\rm rel}=\sqrt{2}b_\Lambda$. The
dashed line incorporates the calculated correlation function,
$\Psi_{^1S_0}(k,r)/j_0(kr)$, a result that is well
parametrized by a phenomenological correlation function of the form of Eq.~~(\ref{eq:fsief}),
as shown by the dotted line.

As mentioned before, the
strong BBM couplings are taken from the NSC97f\cite{nij99}
potential.
In the weak sector only the couplings corresponding to decays
involving pions are experimentally known. The weak couplings
for heavy mesons are obtained following Refs.\cite{holstein,delatorre}. The
PV amplitudes for
the nonleptonic decays $B \rightarrow B' + M$ involving pseudoscalar mesons
are derived using the soft-meson reduction theorem and SU(3) symmetry. This
allows us to
relate the physical amplitudes of the nonleptonic hyperon decays
into a pion plus a nucleon, $B \rightarrow B' + \pi$, with
the unphysical amplitudes involving other members of the meson and
baryon octets.
On the other hand, SU(6$)_W$ permits
relating the amplitudes involving pseudoscalar mesons with those for 
vector mesons.
The weak PC coupling constants are obtained using a pole model,
where the pole terms due to the $(\frac{1}{2})^+$ ground
state (singular in the SU(3) soft meson limit) are assumed to be the dominant
contribution. No meson pole terms are included. The values of
these couplings for the weak vertices 
are listed in Table \ref{tab:wcc}.

\section{Results}

The results for the decay rates of $^6_{\Lambda \Lambda}$He into the
different channels are
summarized in Tables \ref{tab:resfree.nsc97f},\ref{tab:res.nsc97f} and
\ref{tab:diffsi.nsc97f},
where the rate is presented in 
units of the free $\Lambda$ decay, $\Gamma_\Lambda=3.8 \times 10^8~{\rm s}^{-1}$.
Apart from giving the various partial rates, 
the ratio between the neutron-induced and proton-induced
decay rates, $\Gamma_{\rm n}/\Gamma_{\rm p}$, 
is also quoted. Only $\Delta I=1/2$ transitions have been considered.

We begin by discussing the results of Table \ref{tab:resfree.nsc97f}, 
which list calculations absent of
either strong correlations (initial and final) and FF. Although the results are
clearly unrealistic, they serve to illustrate, in comparison with the following table, the
effect of the strong interaction on the weak decay process. 
The $\Lambda N \to NN$ transitions receive contributions mainly from the
$\pi$, $\rho$, $K$ and $K^*$ mesons; however, once short-range strong correlations
are included (see Table \ref{tab:res.nsc97f}), the role of the heavier mesons is
strongly reduced. The results in Table \ref{tab:resfree.nsc97f} illustrate that
isospin symmetry forbids isovector meson  contributions, $(\pi,\rho)$, for the
uncorrelated $\Lambda\Lambda \to \Lambda n$ decay rate and isoscalar meson contributions,
$(\eta,\omega)$, for the uncorrelated $\Lambda \Lambda \to \Sigma N$
rates. Finally, we note that the nucleon-induced rate is about 20 times larger than
the total hyperon-induced contribution.

When FF and strong correlations in the initial and final states are included, the
picture changes considerably, as can be seen from the results shown in Table
\ref{tab:res.nsc97f}.  Most of the $\Lambda N\to NN$ rate comes from the exchange of
the $\pi$ and $K$ mesons. Note that the small $\Gamma_{\rm n}/\Gamma_{\rm p}$ ratio for the
one-pion exchange mechanism substantially increases when the $K$-meson is
incorporated due to a destructive (constructive) interference in the $p$-induced
($n$-induced) rate.  This aspect is discussed recently in Ref.~\cite{PR01}, which updates
the meson-exchange model of Ref.~\cite{PRB97} and
finds agreements with results obtained by groups combining the 
$\pi$ and $K$ exchange with either a direct quark mechanism\cite{SIO00} or
$2\pi$ exchange \cite{JOP01}.  Since there are
now two $\Lambda$ particles contributing to the $\Lambda N \to NN$ decay, one would
expect the result to be roughly twice the rate in $^5_\Lambda$He, which for this potential
turns out to be $0.32 \Gamma_\Lambda$, as shown in Ref.~\cite{PR01}. We obtain
instead a significantly larger value of $\Gamma_{\rm NN}=0.96 \Gamma_\Lambda$.
This finding can be traced to the $\Lambda$ which is more strongly bound in
$^6_{\Lambda \Lambda}$He than it is in $^5_\Lambda$He by about 80\% 
(i.e., the $\Lambda$ single-particle separation energy changes from 3.12 MeV
to around 5 MeV). This increased binding is reflected in the $\Lambda$ single-particle
wave function, with a
harmonic oscillator parameter $b_\Lambda=1.6$ fm instead of the value
$b_\Lambda=1.87$ fm used in the decay of single-$\Lambda$ hypernuclei
\cite{PRB97,PR01}.  As mentioned in Sec. II, the parameter $b_\Lambda=1.6$ fm
simulates the uncorrelated $\Lambda \Lambda$ probability of Ref.~\cite{caro}, which
reproduces the binding energy of the three observed double-$\Lambda$ hypernuclei.
There is also some influence from the slightly different kinematics involved in the
$\Lambda N\to NN$ decay of $^6_{\Lambda\Lambda}$He with respect to that in
$^5_{\Lambda}$He.

From the results presented in Table \ref{tab:res.nsc97f}, 
we observe that the $\Lambda\Lambda \to \Lambda n$ rate receives a tiny
contribution from the isovector
$\pi$ and $\rho$ mesons. This is due to the combined transition
$\Lambda\Lambda \stackrel{\rm weak}{\to} \Sigma {\rm N}
\stackrel{\rm strong}{\to}
\Lambda {\rm N}$, which is possible through the coupling between $\Lambda$N
and 
$\Sigma$N states induced by the strong interaction. Analogously, the
strong $\Lambda {\rm N} - \Sigma$N coupling also
induces very small contributions from the isoscalar mesons to the
$\Sigma$N rate, $\Gamma_{\Sigma {\rm N}}$.

We find the rate $\Gamma_{\Lambda {\rm n}}$ to be dominated by $K$-exchange. We note
that the $\pi +K$ contribution is very similar to the value for 
$K$-exchange alone; adding the $\eta$ meson decreases the rate by
35\%.
However, there is strong evidence that we are overestimating the role of
the $\eta$ meson by using strong $\eta$NN and $\eta \Lambda\Lambda$
couplings that rely on SU(3) symmetry. All indications from
a careful analysis of reactions like $\eta$ photoproduction\cite{cornelius} 
point towards a $\eta$NN coupling much smaller than the one
provided by SU(3) symmetry. In this case, the weak 
decay of double-$\Lambda$ hypernuclei going to final $\Lambda$n states would
be almost solely determined by $K$ exchange and the measurement of this decay
channel would provide valuable information on the weak $\Lambda\Lambda$K 
couplings. In particular, since this partial rate is dominated 
by PC transitions, as can be inferred from the values of the
$\Lambda\Lambda {\rm K}^0$ weak couplings listed in Table \ref{tab:wcc},
one would gain access to the PC $\Lambda\Lambda {\rm K}^0$ amplitude. 
The rate to final $\Sigma$N states, $\Gamma_{\Sigma {\rm N}}$, on the other hand is very small
and, as expected, dominated by the $\pi$-exchange mechanism.

The ratios between the rates that include correlations are:
$$\frac{\Gamma_{\rm NN}}{\Gamma_{\Sigma {\rm N}}}:\frac{\Gamma_{\Lambda {\rm N}}}
{\Gamma_{\Sigma
{\rm N}}}:\frac{\Gamma_{\Sigma {\rm N}}}{\Gamma_{\Sigma {\rm N}}} \simeq 320:12:1 \, .$$
Correlations have a stronger influence on the YN channels, which are a factor
$30-300$ smaller compared to the NN rate,
even though the $\Sigma$N transition allows a final state with 
a lower relative momentum and, therefore, is
less sensitive to the strongly repulsive core of the YN interaction. 
However, the NN rate is dominated by tensor transitions 
that are absent in the YN
channels because the interacting $\Lambda\Lambda$ pair
is in a $^1S_0$ state. With the tensor strength distributed
towards larger distances compared to those for central and parity-violating
transitions, the NN transition becomes comparatively less affected
by strong correlations than the YN ones.
Note that correlations tend to increase the $\Lambda$N rate while decreasing the $\Sigma$N 
rate.
The reason is that the uncorrelated  $\Sigma$N rate is 
dominated by $\pi$ and $K^*$ exchanges, the $K^*$ contribution being about
twice that of the $\pi$ meson, whereas the $\Lambda$N rate comes
mainly from strange meson exchange, with the $K$ contribution  
twice as large as that of the $K^*$. Therefore, the supression of short-range physics
induced by correlations affects the $\Sigma$N rate more than the $\Lambda$N one.
In addition, it is known\cite{VPR00,JD99} that the $\Sigma$N 
interaction  for the new Nijmegen potentials is quite 
repulsive in the $I=1/2$ channel, affecting the $\Sigma$N wave
functions significantly, which are pushed out from the weak potential interaction
range. 

The results for various approaches to the treatment of FSI are compared in
Table \ref{tab:diffsi.nsc97f}. Those labelled with ``FSI eff."
have been obtained by multiplying the free NN, $\Lambda$N and $\Sigma$N
final states by the momentum and channel independent 
function \mbox{$1-j_0(q_c r)$}, with $q_c=3.93$
fm$^{-1}$. 
As recently discussed in Ref.~\cite{PR01}, treating FSI in this simplified
phenomenological fashion overestimates the rate
$\Gamma_{\rm NN}$ calculated rigorously with a $T$-matrix by about a factor of two. 
The rate $\Gamma_{\Lambda {\rm n}}$ obtained with the phenomenological FSI
treatment is almost one order of magnitude smaller compared with the result obtained within a
$T$-matrix approach, while on the other hand, the rate $\Gamma_{\Sigma {\rm N}}$ is
overpredicted by almost an order of magnitude.
As in Ref.~\cite{PR01}, we again find that in
 order to be able to extract information on the weak decay amplitudes 
a careful treatment of FSI is essential in
the study of the weak decay of double-$\Lambda$ hypernuclei.

We finally compare our results with the ones presented recently
by Itonaga et al. \cite{ito01}, where the weak decay mechanism contains $\pi$, $\rho$ 
and correlated two-pion exchange with scalar ($\sigma$) and vector
($\rho$) quantum numbers. 
The $\Lambda$ wave function in $^6_{\Lambda \Lambda}$He was taken the same
as in their $^5_\Lambda$He calculation \cite{ito95},
thus neglecting the possible effect of a stronger binding of
the $\Lambda$ in a double-$\Lambda$ hypernucleus.
Hence, the nucleon-induced rate obtained in \cite{ito01} is
$\Gamma_{\rm NN}=0.753 \Gamma_\Lambda$, equal to twice their decay rate in
$^5_\Lambda$He. This rate is about 20\%
smaller than the one obtained here. The rates for $\Lambda$n and 
$\Sigma {\rm N}$ final states, $\Gamma_{\Lambda {\rm n}}=0.025 \Gamma_\Lambda$ and  
$\Gamma_{\Sigma {\rm N}}=0.0012  \Gamma_\Lambda$, are about a factor two
smaller 
than the ones obtained in this work. We note that the $\Lambda \Lambda$ 
correlation function used in Ref. \cite{ito01} is in fact the $^1S_0$
$\Lambda$N one used in their study of the decay of 
$^5_\Lambda$He.
We also note that the
$K$-exchange mechanism, dominant in the $\Lambda \Lambda \to
\Lambda n$ transition amplitude, is absent in the work of
Ref.~\cite{ito01}. 

In Ref. \cite{PRB97}, we explored SU(3) symmetry breaking effects in the decay of
single-$\Lambda$ hypernuclei by using weak NNK couplings
derived in the framework of heavy baryon chiral perturbation theory\cite{SS96}, 
where one-loop SU(3) corrections to leading order were evaluated. 
Here we explore similar effects not only for the NNK but also for
 the $\Lambda \Lambda$K and $\Lambda \Sigma$K couplings,
which were calculated in Refs. \cite{roxanne,SR00}. In \mbox{Table \ref{ChPT}} we
list the values 
for the weak PV ($S$-wave) and PC ($P$-wave) couplings involving the kaon. 
Compared to the ones quoted in Ref. \cite{PRB97}, the numbers in Table \ref{ChPT} differ
since they were obtained using
a common mass splitting of $200$ MeV between the octet and decuplet 
baryons \cite{roxanne,SR00}.

Comparing the values listed in Table \ref{ChPT} with those in Table \ref{tab:wcc}
we  observe that the one-loop corrected NNK
$P$-wave constants are approximately a factor two smaller
than the tree level ones, while the $S$-wave are very similar, except for the 
$pn K^+$ coupling, which, again, is half the tree-level value.
This explains the reduced $K$-exchange contribution to the nucleon-induced rates
with respect to the tree-level result.
Note that the magnitude of the (constructive) 
interference between $\pi$ and $K$ in the $nn$ channel has not changed much, 
while it has become less destructive in the $np$ channel, thereby producing a smaller
$\Gamma_{\rm n}/\Gamma_{\rm p}$ ratio. 
Once all the mesons are included, the NN rate obtained with the
one-loop corrected NNK couplings is a factor 1.4 larger than that obtained with
the tree-level values, while the ratio $\Gamma_{\rm n}/\Gamma_{\rm p}$ is smaller by a 
factor of
two.

Although the one-loop corrected $\Lambda \Lambda$K couplings of Table \ref{ChPT}
are not much different from the corresponding tree-level values, they lead to a
reduction of the $K$-meson exchange contribution to the
$\Lambda$n rate by about a factor two. Hence, the contribution of the
$\eta$ meson and the vector mesons becomes more relevant, making the
extraction of the $\Lambda\Lambda$K coupling from this $\Lambda \Lambda \to \Lambda
n$ decay mode more difficult.

The decay rate into final $\Sigma$N states is the only one that increases when
the one-loop corrected $\Lambda \Sigma$K couplings are used, despite the magnitude of the 
$P$-wave amplitudes
being ten times smaller than the
corresponding tree-level values, whereas the $S$-wave amplitudes are only 
moderately larger. In fact, Table 
\ref{tab:res.nsc97f} reveals that the $K$-exchange contribution to the $\Sigma$N rate is
two times smaller than that obtained with the tree-level $\Lambda \Sigma$K
couplings, as expected from the reduced magnitude of the couplings. 
However, due to the opposite sign of the 
loop-corrected $P$-wave coupling with respect to the corresponding tree level
value,
the interference between the $\pi$ and $K$ contributions is now constructive
instead of destructive,
which explains the final increased loop-corrected rate.

Finally, we have investigated how our results would change if we allowed the
one-loop corrected BBK couplings to vary within their estimated error bands. The
corresponding range of variation on the rates is quoted in Table
\ref{tab:summarize}, where the results obtained with the tree-level BBK couplings
have also been included to facilitate the comparison between both calculations.
The results summarized in Table \ref{tab:summarize} show clearly that the inclusion
of weak BBK couplings obtained with one-loop
corrections to the leading order in $\chi$PT produces
a $\Lambda N \to NN$ rate 40 \% larger than the tree-level value,
a ratio $\Gamma_{\rm n}/\Gamma_{\rm p}$ twice smaller,
a $\Lambda \Lambda \to \Lambda n$ rate roughly twice smaller, 
and a $\Lambda \Lambda \to \Sigma N$ rate almost twice as large.

\section{Conclusions}

In this study we investigated the nonmesonic weak decay modes of 
double-$\Lambda$ hypernuclei within a one-meson-exchange framework. 
Our results are given for the specific double-$\Lambda$
hypernucleus $^6_{\Lambda \Lambda}$He, but can easily be extended to heavier 
systems. 

The standard nucleon-induced nonmesonic $\Lambda N \to NN$ rate
is found to be more than twice as large as in $^5_{\Lambda}$He due to 
the increased binding of the second $\Lambda$ hyperon. 
Two new hyperon-induced modes become possible,
$\Lambda\Lambda \to \Lambda n$ and $\Lambda\Lambda \to \Sigma N$, the latter
coming in two charge states.  We find the total hyperon-induced rate to be as 
large as 4\% of the total nonmesonic rate. 
This new rate is dominated by the $\Lambda\Lambda \to \Lambda n$ mode. 
In fact, this transition turns out to be the more interesting one 
(rather than the $\Lambda\Lambda \to \Sigma N$) since it allows direct access 
to exotic vertices like $\Lambda \Lambda$K, unencumbered by the usually 
dominant pion exchange. Indeed,
one-loop log corrected $\chi$PT results modify the $\Lambda\Lambda \to 
\Lambda n$ by 50\% while changing the $\Lambda N \to NN$ only at the 15\% level,
demonstrating the power of this weak mechanism to test $\chi$PT in the weak 
SU(3) sector.  With a free $\Lambda$ in the final state
this new mode should be distinguishable from the usual nucleon-induced decay
channels.  Given the potential benefits to hadronic physics,
the experimental program of investigating the production and decay
of double-$\Lambda$ hypernuclei should be intensified.

\section*{Acknowledgements}
The authors would like to thank Christoph Hanhart, Barry R. Holstein, Roxanne
P. Springer and Isaac Vida\~na for their respective inputs and comments to different
aspects of the present paper, regarding both the strong and weak baryon-baryon 
interaction. One of the authors, A.P., would like to thank the Institute for Nuclear 
Theory and the University of Washington 
(Seattle, USA) where an important part of this work was developped.
This work has been partially supported by the U.S. Dept. of Energy under 
Grant No. DE-FG03-00-ER41132, by the DGICYT (Spain) under contract
PB98-1247 and by the Generalitat de Catalunya project SGR2000-24.

\section{Appendix} 

\subsection{Weak PC couplings} 

In this subsection we present the expression for the PC weak coupling
constants derived by using a pole model\cite{holstein} with only baryon poles.

\begin{itemize}

\item{\bf $\pi$-exchange}
\begin{eqnarray} {\cal A}(\Lambda,n \pi^0) &=&
g(n,n\pi^0) \frac{A_{\Lambda n}}{m_\Lambda - m_n} + g(\Lambda, \Sigma^0 \pi^0)
\frac{A_{\Sigma^0 n}}{m_n - m_{\Sigma^0}} \, \ ,
\label{eq:wpcpion}
\\ {\cal A}(\Lambda,p \pi^-) &=& g(n,p
\pi^-) \frac{A_{\Lambda n}}{m_\Lambda - m_n} + g(\Lambda, \Sigma^+ \pi^-)
\frac{A_{\Sigma^+ p}}{m_p - m_{\Sigma^+}} \, \ {\rm .} 
\end{eqnarray} 

\item{\bf $\eta$-exchange}
\begin{eqnarray}
{\cal A}(\Lambda,n \eta)&=& g(n,n \eta) \frac{A_{\Lambda n}}{m_\Lambda - m_n} +
g(\Lambda, \Lambda \eta) \frac{A_{\Lambda n}}{m_n - m_\Lambda} \, \ {\rm .}  
\label{eq:wpceta}
\end{eqnarray}

\item{\bf $K$-exchange}
\begin{eqnarray}
{\cal A}(n, n K^0) &=& g(n, \Lambda K^0) \frac{A_{\Lambda n}}{m_n - m_\Lambda} +
g(n, \Sigma^0 K^0) \frac{A_{\Sigma^0 n}}{m_n - m_{\Sigma^0}}\, \ , \\
{\cal A}(p, p K^0) &=& g(p, \Sigma^+ K^0) \frac{A_{\Sigma^+ p}}{m_p - 
m_{\Sigma^+}} \, \ ,\\
{\cal A}(p, n K^+) &=& g(p, \Sigma^0 K^+) \frac{A_{\Sigma^0 n}}
{m_n - m_{\Sigma^0}} +
g(p, \Lambda K^+) \frac{A_{\Lambda n}}{m_n - m_\Lambda} \, \ ,\\
{\cal A}(\Lambda, \Lambda K^0) &=& g(n, \Lambda K^0) \frac{A_{\Lambda n}}
{m_\Lambda - m_n} + g(\Lambda, \Xi^0 K^0) \frac{A_{\Xi^0 \Lambda}}{m_\Lambda - 
m_\Xi^0}\, \ ,\\
{\cal A}(\Lambda, \Sigma^0 K^0) &=& g(n, \Sigma^0 K^0)
\frac{A_{\Lambda n}}{m_\Lambda - m_n} \, \ ,\\
{\cal A}(\Lambda, \Sigma^- K^+) &=& g(n, \Sigma^- K^+) 
\frac{A_{\Lambda n}}{m_\Lambda - m_n} \, \ {\rm .}
\label{eq:wpckaon}
\end{eqnarray}
  
\end{itemize}

To compute the former amplitudes, we use
the following values for the weak
transitions at the baryon lines $A_{\rm BB'}$:

\begin{eqnarray}
A_{\Lambda n} &\equiv& A_{\Lambda {\rm N}} = - 4.32 \times 10^{-5} {\rm MeV} \, \ ,
\nonumber \\
A_{\Sigma^0 n} &\equiv& - A_{\Sigma {\rm N}} = 4.35 \times 10^{-5} {\rm MeV}\, \ ,
\nonumber \\ 
A_{\Xi^0 \Lambda} &\equiv& A_{\Xi \Lambda} = 5.93 \times 10^{-5} {\rm MeV} \, \ ,
\nonumber \\
A_{\Sigma^+ p} &\equiv& \sqrt{2} A_{\Sigma {\rm N}} = - 6.15 \times 10^{-5} {\rm MeV} \, \ {\rm .}
\end{eqnarray}

For the vector $\rho$, $\omega$ and ${K^*}$ mesons one has to make the following
replacements in Eqs. (\ref{eq:wpcpion}) to (\ref{eq:wpckaon}):
\begin{eqnarray}
\pi &\to& \rho, \nonumber \\
\eta &\to& \omega, \nonumber \\
K &\to& {K^*}.
\label{vmrep}
\end{eqnarray}

\subsection{Weak PV couplings}

SU(3) allows us to obtain the weak PV couplings involving two baryons and one 
pseudoscalar ($\pi, \eta, K$) meson. 
In order to obtain the corresponding constants for vector mesons we use
SU(6)$_W$. Here, we quote the results for all the couplings that appear in
the $\Lambda N \to NN$ and $\Lambda N \to YN$ channels.

\begin{eqnarray}
(\Lambda,p \pi^-)_{\rm exp} &\equiv& \Lambda ^0_- = 3.25 \times 10^{-7} \, \ ,\\
(\Sigma ^+, p \pi^0)_{\rm exp} &\equiv& \Sigma ^+_0 =  -3.27 \times 10^{-7}\, \ ,  \\
(\Lambda,n \pi^0) &=& - \frac{1}{\sqrt{2}} \Lambda ^0_-\, \ , \\
(\Lambda,n \rho^0) &=& \frac{1}{3} (-2 \Lambda ^0_- + \sqrt{3} \Sigma ^+_0 -
\sqrt{3} a_T) \, \ ,\\
(\Lambda,p \rho^-) &=& \frac{1}{3} \sqrt{\frac{2}{3}} (2 \sqrt{3} \Lambda ^0_- -
3  \Sigma ^+_0  -3 a_V )\, \ , \\
(\Lambda,n \eta) &=& \sqrt{\frac{3}{2}} \Lambda ^0_-\, \ , \\
(\Lambda,n \omega) &=& \Sigma ^+_0 - \frac{1}{3} a_T\, \ , \\  
(n, n K^0)&=& \sqrt{\frac{3}{2}} \Lambda ^0_- - \frac{1}{\sqrt{2}} \Sigma ^+_0 \, \ ,\\
(p, p K^0)&=& - \sqrt{2} \Sigma ^+_0 \, \ ,\\
(p, n K^+)&=& \sqrt{\frac{3}{2}} \Lambda ^0_- + \frac{1}{\sqrt{2}} \Sigma ^+_0 \, \ ,\\
(n, n K^{*^0}) &=& - \sqrt{3}\Lambda ^0_- - \frac{1}{3} \Sigma ^+_0 - 
\frac{2}{9} a_T\, \ , \\
(p, p K^{*^0}) &=& - \frac{2}{3} \Sigma ^+_0 + \frac{8}{9} a_T\, \ , \\
(p, n K^{*^+}) &=& - \sqrt{3}\Lambda ^0_- + \frac{1}{3} \Sigma ^+_0 + 
\frac{10}{9} a_V\, \ , \\
(\Lambda, \Lambda K^0) &=& - \frac{1}{2}\sqrt{\frac{3}{2}} \Lambda ^0_- -
\frac{3}{2 \sqrt{2}} \Sigma ^+_0 \, \ , \\ 
(\Lambda, \Sigma^0 K^0)&=& - \frac{1}{2 \sqrt{2}} \Lambda ^0_- - \frac{1}{2} 
\sqrt{\frac{3}{2}} \Sigma ^+_0 \, \ , \\
(\Lambda, \Sigma^- K^+)&=& \frac{1}{2} \Lambda ^0_- + \frac{\sqrt{3}}{2}
 \Sigma ^+_0 \, \ , \\
(\Lambda, \Lambda K^{*^0}) &=&  - \frac{5 \sqrt{3}}{6} \Lambda ^0_- - \frac{1}{2} 
\Sigma ^+_0\, \ , \\ 
(\Lambda, \Sigma^0 K^{*^0}) &=& \frac{7}{6} \Lambda ^0_- + \frac{3}{2 \sqrt{3}} 
\Sigma ^+_0 + \frac{2}{3 \sqrt{3}} a_T\, \ , \\ 
(\Lambda, \Sigma^- K^{*^+}) &=& - \frac{7}{3 \sqrt{2}} \Lambda ^0_- - \frac{3}{\sqrt{6}} 
\Sigma ^+_0 + \frac{4}{3 \sqrt{6}} a_V  \, \ {\rm .}
\end{eqnarray}

If we desire only the $\Delta I= \frac{1}{2}$ part of the
above expressions, as we do in the present work, the following replacements 
have to be made\cite{holstein.private}:

\begin{itemize}

\item For the $\Lambda {\rm N} \rho$ couplings:
$a_V \to 3 a_V $
and 
$a_T \to 3 a_T$.

\item For the $\Lambda \Sigma {\rm K}^*$ couplings:
$a_V \to \displaystyle\frac{a_V - a_T}{\sqrt{3}}$
and
$a_T \to - \displaystyle\frac{a_V - a_T}{\sqrt{3}}$.  

\end{itemize}
 
\subsection{Strong coupling constants}

In this subsection we present the convention used for the strong couplings.

\begin{eqnarray}
g_{{\rm NN} \pi} &\equiv& g(p,p\pi^0) = 
- g(n,n\pi^0) = \frac{1}{\sqrt{2}} g(p,n \pi^+) =
\frac{1}{\sqrt{2}} g(n,p \pi^-)\, \ , \\
g_{\Lambda \Sigma \pi} &\equiv& g(\Sigma^+, \Lambda \pi^+) = 
g(\Sigma^0, \Lambda \pi^0) = g(\Sigma^-, \Lambda \pi^-) = g(\Lambda,
\Sigma^+ \pi^-) \nonumber
\\ &=& 
g(\Lambda, \Sigma^0 \pi^0) = g(\Lambda, \Sigma^- \pi^+)\, \ , \\
g_{N \Lambda K} &\equiv& g(p,\Lambda K^+) = g(n,\Lambda K^0) = g(\Lambda,p K^-) 
= g(\Lambda,n {\bar{K^0}})\, \ , \\
g_{N \Sigma K} &\equiv& g(p, \Sigma^0 K^+) = - g(n,\Sigma^0 K^0) = g(\Sigma^0, pK^-)
= - g(\Sigma^0, n {\bar{K^0}}) \nonumber \\
&=& \frac{1}{\sqrt{2}} g(p, \Sigma^+ K^0) = \frac{1}{\sqrt{2}} 
g(n, \Sigma^- K^+) = \frac{1}{\sqrt{2}} g(\Sigma^-, n K^-) \nonumber \\
&=&
\frac{1}{\sqrt{2}} g(\Sigma^+, p {\bar{K^0}})\, \ ,\\
g_{\Lambda \Xi K} &\equiv& g(\Lambda, \Xi^0 K^0) = g(\Xi^0, \Lambda {\bar K}^0)=
- g(\Lambda, \Xi^- K^+) = - g(\Xi^-, \Lambda K^-)\, \ , \\
g_{{\rm NN} \eta} &\equiv& g(n,n \eta) = g(p, p\eta) \, \ {\rm .} 
\end{eqnarray}

For the vector $\rho$, $\omega$ and ${K^*}$ mesons, the expressions are
equivalent to the ones quoted above by making the replacements
of Eq.~(\ref{vmrep}).

\begin{table}
\caption{Pseudoscalar (PS) and vector (V) hamiltonians entering 
Eq.~(\ref{eq:uncF0}) (in units
of $G_F {m_\pi}^2=2.21 \times 10^{-7}$). $\Psi$ ($\Phi$) 
stands for the baryon (meson) 
field.}\bigskip
\begin{tabular}{lcc}
 & PS & V \\
\hline\\
Strong &  $ {\rm i} g {\overline{\Psi}} \gamma_5 \phi \Psi$ &
$ {\overline{\Psi}} \left[ g^{\rm V} \gamma^\mu + {\rm i}
\displaystyle\frac{ g^{\rm T}}{2M}\sigma^{\mu \nu} q_\nu \right] \phi \Psi$ \\
 & & \\
\hline\\
 & & \\
Weak & ${\rm i} {\overline{\Psi}} (A + B \gamma_5) \phi \Psi$ &
$ {\overline{\Psi}} \left[\alpha \gamma^\mu - \beta {\rm i}
\displaystyle\frac{\sigma^{\mu \nu} q_\nu}
{2 \overline{M}}+\varepsilon\gamma^\mu \gamma_5 \right] \phi \Psi$ \\
 & & \\
\end{tabular}
\label{tab:vert}
\end{table}

\begin{table}
\caption{Possible $^{2S+1}L_J$ channels present in the weak decay
of $_{\Lambda \Lambda}^6$He. The notation used is the following: 
$Lr$ stands for the relative orbital
momentum for the initial $\Lambda \Lambda$ or $\Lambda$N pair, 
$L'$ for the relative angular momentum of the two-particle state right 
after the weak transition occurs, and $L$ for the relative angular momentum 
of the final two-particle state after including the effects of the 
strong interaction (FSI). The symbols C, T and PV stand for Central,
Tensor and 
Parity Violating.  
}\bigskip
\begin{tabular}{|l|ccc|ccc|ccc|}
& \multicolumn{3}{c|}{$\Lambda p \to np$} & 
\multicolumn{3}{c|}{$\Lambda n \to nn$} &
\multicolumn{3}{c|}{$\Lambda \Lambda \to YN$} \\
\hline
&  $L_r$ & $L'$ & $L$ &
 $L_r$ & $L'$ & $L$ &
 $L_r$ & $L'$ & $L$ 
 \\
\hline
C & $^1S_0$ 
$\to$ 
& $^1S_0$ $\to$ & $^1S_0$ & 
$^1S_0$ $\to$ & $^1S_0$ $\to$ & $^1S_0$ &
$^1S_0$ $\to$ & $^1S_0$ $\to$ & $^1S_0$ \\
  & $^3S_1$ $\to$ & $^3S_1$ $\to$ & $^3S_1$, $^3D_1$ &  & & & & & \\
\hline
T & $^3S_1$ $\to$ & $^3D_1$ $\to$ & $^3D_1$, $^3S_1$ &  & & & & & \\
\hline
PV & $^1S_0$ $\to$ & $^3P_0$ $\to$ & $^3P_0$ &
$^1S_0$ $\to$ & $^3P_0$ $\to$ & $^3P_0$ &
$^1S_0$ $\to$ & $^3P_0$ $\to$ & $^3P_0$ \\
   & $^3S_1$ $\to$ & $^1P_1$ $\to$ & $^1P_1$  & & & & & & \\
   & $^3S_1$ $\to$ & $^3P_1$ $\to$ & $^3P_1$ & 
$^3S_1$ $\to$ & $^3P_1$ $\to$ & $^3P_1$ &  & & \\
\end{tabular}
\label{tab:chan}
\end{table}

\begin{table}
\caption{Cut-off values (in MeV) used in the present calculation for a FF of
the type
$\widetilde{\Lambda}_i^2/(\widetilde{\Lambda}_i^2 + \vec{q}\,^2)$, which
matches the exponential-type FF used in Ref.~\protect\cite{nij99}}. \bigskip
\begin{tabular}{|c|c|c|c|c|c|}                          
$\pi$ & $\eta$ & $K$ & $\rho$ & $\omega$ & $K^*$ \\
\hline
1750 & 1750 & 1789 & 1232 & 1310 & 1649 \\
\end{tabular}
\label{tab:ff}
\end{table}         

\begin{table}
\caption{Parity Conserving (PC) and Parity Violating (PV) 
weak coupling constants 
for the decay of $^6_{\Lambda \Lambda}$He. The numbers are in units of 
$G_F {m_\pi}^2 = 2.21 \times 10^{-7}$. The NSC97f\protect\cite{nij99} model has
been used for the strong sector.}\bigskip
\begin{center}
\begin{minipage}{8cm}
\begin{tabular}{|c|c|c|}
 & {PV ($S$-wave)} & {PC ($P$-wave)}   \\
\hline
$\Lambda n \pi^0$ & $-1.05$ & 7.15 \\
\hline
$\Lambda p \pi^-$ & 1.48 & $-10.11$ \\
\hline
$\Lambda n \eta$ & 1.80 & $-11.90$  \\
\hline
$pn K^+$ & 0.76 & $-23.70$ \\
\hline
$pp K^0$ & 2.09 & 8.33  \\
\hline
$nn K^0$ & 2.85 & $-15.37$ \\
\hline
$\Lambda \Lambda {\rm K}^0$ & 0.67 & 12.72  \\
\hline
$\Lambda \Sigma^0 {\rm K}^0$ & 0.39 & 5.95  \\
\hline
$\Lambda \Sigma^- {\rm K}^+$ & $-0.55$ & $-8.42$ \\
\hline
$(\Lambda n \rho^0)$ & $-1.09$ &  (V) $3.29$  \\
 & &  (T) 6.74   \\
\hline
$(\Lambda p \rho^-)$ & 1.54 & (V) $-4.65$  \\
&  & (T) $-9.53$   \\
\hline
$(\Lambda n \omega)$ & $-1.33$ & (V) $-0.17$  \\
 & &  (T) $-7.43$  \\
\hline
$(pp {K^*}^0)$ & 0.60 & (V) $-5.46$ \\
 & & (T) 6.23  \\
\hline
$(pn {K^*}^+)$ & $-4.48$ &  (V) $-4.02$ \\
 & & (T) $-19.54$  \\
\hline
$(nn {K^*}^0)$ &$-3.88$ &  (V) $-9.48$ \\
 & & (T) $-13.31$  \\
\hline
$(\Lambda \Lambda {K^*}^0)$ & $-1.38$ & (V) $-1.34$ \\
&  &  (T) 11.20 \\
\hline
$(\Lambda \Sigma^0 {{\rm K}^*}^0)$ & 0.63 & (V) $-3.90$ \\
&  & (T) 4.45  \\
\hline
$(\Lambda \Sigma^- {{\rm K}^*}^+)$ & $-0.88$ & (V) $5.51$ \\
& & (T) $-6.30$ \\
\end{tabular}
\end{minipage}
\end{center}
\label{tab:wcc}
\end{table}

\newpage

\begin{table}
\caption{Individual and combined meson-exchange contributions to the 
nonmesonic weak decay
of $^6_{\Lambda \Lambda}$He in absence of
either strong correlations and FF. The nonmesonic
decay rate is in units of
$\Gamma_\Lambda = 3.8 \times 10^{9} {\rm s}^{-1}$.
The NSC97f~\protect\cite{nij99} strong coupling constants
have been used.}
\bigskip                         
\begin{tabular}{|l|c|c|c|c|c|c|c|}
& \multicolumn{3}{c|}{$\Gamma_{\rm NN}$} & $\Gamma_{\Lambda {\rm n}}$ &
\multicolumn{2}{c|}{$\Gamma_{\Sigma {\rm N}}$} & $\Gamma_{\rm NN} + \Gamma_{\rm YN}$ \\
\hline
Meson & $\Gamma_{\rm nn}$ & $\Gamma_{\rm np}$ & 
$\Gamma_{\rm n}/\Gamma_{\rm p}$ &
& $\Gamma_{\Sigma^0 {\rm n}}$ &
$\Gamma_{\Sigma^- {\rm p}}$ &   \\
\hline
$\pi$ & 0.40 & 2.08 & 0.19 & $--$ & 2.1 $\times 10^{-2}$ & 
4.2 $\times 10^{-2}$ & 2.54 \\
\hline
$K$ & 0.67 & 1.14 & 0.58 & 3.5 $\times 10^{-2}$ & 2.3 $\times 10^{-3}$
& 4.6 $\times 10^{-3}$ & 1.85 \\ 
\hline
$\eta$ & 3.5 $\times 10^{-2}$ & 3.0 $\times 10^{-2}$ & 1.18 &
2.0 $\times 10^{-3}$ 
& $--$ & $--$ & 6.6 $\times 10^{-2}$ \\
\hline
$\rho$ & 0.30 & 0.70 & 0.42 & $--$ & 2.2 $\times 10^{-3}$ & 
4.5 $\times 10^{-3}$ & 1.00 \\
\hline
$K^*$ & 3.46 & 2.31 & 1.50 & 1.6 $\times 10^{-2}$ & 3.8 $\times 10^{-2}$ & 
7.7 $\times 10^{-2}$ & 5.90 \\
\hline 
$\omega$ & 8.6 $\times 10^{-2}$ & 6.6 $\times 10^{-2}$ & 1.31 & 
6.8 $\times 10^{-3}$ &$--$ & $--$ & 0.16 \\
\hline
\hline
$\pi + K$ & 0.96 & 1.09 & 0.88 & 3.5 $\times 10^{-2}$ & 1.0 $\times 10^{-2}$ &
2.0 $\times 10^{-2}$ & 2.11 \\
\hline
$\pi + K + \eta$ & 1.19 & 1.06 & 1.12 & 2.6 $\times 10^{-2}$ & 
1.0 $\times 10^{-2}$ & 2.0 $\times 10^{-2}$ & 2.31 \\ 
\hline
\hline
{\rm all} & 2.45 & 4.29 & 0.57 & 0.10 & 8.7 $\times 10^{-2}$ & 0.17 & 7.11 \\
\end{tabular}
\label{tab:resfree.nsc97f}
\end{table}

\begin{table}
\caption{
Individual and combined meson-exchange contributions to the nonmesonic 
weak decay
of $^6_{\Lambda \Lambda}$He. Strong correlations (initial and final)
and FF are included.
The nonmesonic
decay rate is in units of
$\Gamma_\Lambda = 3.8 \times 10^{9} {\rm s}^{-1}$.
The NSC97f~\protect\cite{nij99} strong interaction model
has been used.
The values between parenthesis have been obtained using the weak kaon   
couplings from Refs. \protect\cite{roxanne,SR00}}. \bigskip
\begin{tabular}{|l|c|c|c|c|c|c|c|}
& \multicolumn{3}{c|}{$\Gamma_{\rm NN}$} & $\Gamma_{\Lambda {\rm n}}$ &
\multicolumn{2}{c|}{$\Gamma_{\Sigma {\rm N}}$} & $\Gamma_{\rm NN} + \Gamma_{\rm YN}$ \\
\hline
Meson & $\Gamma_{\rm nn}$ & $\Gamma_{\rm np}$ &
$\Gamma_{\rm n}/\Gamma_{\rm p}$ &   & $\Gamma_{\Sigma^0 {\rm n}}$ &
$\Gamma_{\Sigma^- {\rm p}}$ &   \\
\hline
$\pi$ & 9.9 $\times 10^{-2}$ & 1.09 & 9.1 $\times 10^{-2}$ & 
1.3 $\times 10^{-4}$ & 1.8 $\times 10^{-3}$ & 3.7 $\times 10^{-3}$ & 1.19 \\
\hline
$K$ & 8.2 $\times 10^{-2}$ & 0.47 & 0.17 & 
2.7 $\times 10^{-2}$ & 3.1 $\times 10^{-4}$ & 6.1 $\times 10^{-4}$ & 
0.58  \\ 
& (5.3 $\times 10^{-2}$) & (0.11) & (0.50) & (1.3 $\times 10^{-2}$) &
(1.7 $\times 10^{-5}$) & (3.4 $\times 10^{-5}$) & (0.17) \\
\hline
$\eta$ & 3.9 $\times 10^{-3}$ & 7.6 $\times 10^{-3}$ & 0.51 & 
1.1 $\times 10^{-3}$ & 2.5 $\times 10^{-7}$ & 5.1 $\times 10^{-7}$ & 
1.3 $\times 10^{-2}$ \\
\hline
$\rho$ & 7.4 $\times 10^{-3}$ & 2.5 $\times 10^{-2}$ & 0.30 & 
4.7 $\times 10^{-6}$ & 2.0 $\times 10^{-6}$ & 4.0 $\times 10^{-6}$ &
3.2 $\times 10^{-2}$ \\
\hline
$K^*$ & 3.8 $\times 10^{-3}$ & 2.5 $\times 10^{-2}$ & 0.15 & 
4.8 $\times 10^{-3}$ & 7.3 $\times 10^{-5}$ & 1.5 $\times 10^{-4}$ & 
3.4 $\times 10^{-2}$ \\
\hline
$\omega$ &  1.1 $\times 10^{-3}$ & 0.95 $\times 10^{-3}$ & 0.12 & 
7.2 $\times 10^{-5}$ & 2.5 $\times 10^{-8}$ & 5.0 $\times 10^{-8}$ & 
2.1 $\times 10^{-3}$  \\
\hline
\hline
$\pi + K$ & 0.24 & 0.48 & 0.50 & 2.6 $\times 10^{-2}$ & 1.2 $\times 10^{-3}$ & 
2.3 $\times 10^{-3}$ & 0.75 \\
 & (0.21) & (0.82) & (0.25) & (1.3 $\times 10^{-2}$) & 
(2.2 $\times 10^{-3}$) & (4.4 $\times 10^{-3}$) & (1.05) \\ 
\hline
$\pi + K + \eta$ & 0.27 & 0.44 & 0.61 & 1.7 $\times 10^{-2}$ & 
1.1 $\times 10^{-3}$ & 2.3 $\times 10^{-3}$ & 0.73 \\
 & (0.24) & (0.73) & (0.33) & (6.6 $\times 10^{-3}$) &
(2.1 $\times 10^{-3}$) & (4.3 $\times 10^{-3}$) & (0.99) \\ 
\hline
\hline                                                             
{\rm all} & 0.30 & 0.66 & 0.46 & 3.58 $\times 10^{-2}$ 
& 1.0 $\times 10^{-3}$ 
& 2.0 $\times 10^{-3}$ & 1.00 \\
 & (0.25) & (1.13) & (0.23) & (1.9 $\times 10^{-2}$) & 
(1.7 $\times 10^{-3}$) & (3.4 $\times 10^{-3}$) & (1.40) \\   
\end{tabular}
\label{tab:res.nsc97f}
\end{table}

\begin{table}
\caption{Decay rate of $^6_{\Lambda \Lambda}$He in units of 
$\Gamma_\Lambda = 3.8 \times 10^{9} {\rm s}^{-1}$ for different 
approaches to FSI. All mesons are included in the
calculation.}\bigskip
\begin{tabular}{|l|c|c|c|c|c|c|c|}
& \multicolumn{3}{c|}{$\Gamma_{\rm NN}$} & $\Gamma_{\Lambda {\rm n}}$ &
\multicolumn{2}{c|}{$\Gamma_{\Sigma {\rm N}}$} & $\Gamma_{\rm NN} + \Gamma_{\rm YN}$ \\
\hline
 & $\Gamma_{\rm nn}$ & $\Gamma_{\rm np}$ &
$\Gamma_{\rm n}/\Gamma_{\rm p}$ &  & $\Gamma_{\Sigma^0 {\rm n}}$ &
$\Gamma_{\Sigma^- {\rm p}}$ &   \\
\hline
NO FSI & 0.69 & 1.21 & 0.57 & 2.5 $\times 10^{-3}$ & 1.1 $\times 10^{-2}$ &
2.2 $\times 10^{-2}$ & 1.93 \\
\hline
FSI eff. & 0.73 & 1.30 & 0.58 & 4.4 $\times 10^{-3}$ & 8.0 $\times 10^{-3}$ &
1.6 $\times 10^{-2}$ & 2.03 \\
\hline
T-matrix & 0.30 & 0.66 & 0.46 & 3.6 $\times 10^{-2}$ & 1.0 $\times 10^{-3}$ & 
2.0 $\times 10^{-3}$ & 1.00 \\
\end{tabular}
\label{tab:diffsi.nsc97f}
\end{table} 

\begin{table}
\caption{$\chi$PT one-loop corrected NNK, $\Lambda \Lambda$K and
$\Lambda \Sigma$K couplings. Values taken from 
Refs.~\protect\cite{roxanne,SR00}.}\bigskip
\begin{tabular}{|l|c|c|}
\ & $S$-wave & $P$-wave \\
\hline
${\rm pnK}^+$ & $(0.32 \pm 0.24)$ & $(-10.41 \pm 1.61)$ \\
${\rm ppK}^0$ & $(2.32 \pm 0.22)$ & $(3.47 \pm 2.23)$ \\
${\rm nnK}^0$ & $(2.64 \pm 0.25)$ & $(-6.94 \pm 2.23)$ \\
\hline
$\Lambda \Lambda {\rm K}^0$ & $(0.78 \pm 0.20)$ & $(8.69 \pm 2.21)$ \\
\hline
$\Lambda \Sigma^0 {\rm K}^0$ & $(0.53 \pm 0.12)$ & $(-0.58 \pm 1.83)$ \\
$\Lambda \Sigma^- {\rm K}^+$ & $(-0.75 \pm 0.17)$ & $(0.82 \pm 2.59)$ \\
\end{tabular}
\label{ChPT}
\end{table}

\begin{table}
\caption{Partial weak decay rates for $^6_{\Lambda \Lambda}$He
(in units of
$\Gamma_\Lambda = 3.8 \times 10^{9} {\rm s}^{-1}$).
}
\begin{tabular}{|l|c|c|}
\hline
 & NSC97f  & NSC97f + one-loop corrections \\
\hline
$\Lambda n \to nn$ & 0.30 & 0.25 $\pm$ 15 \%\\
$\Lambda p \to np$ & 0.66 & 1.13 $\pm$ 15 \% \\
\hline
$\Lambda N \to NN$ & 0.96 & 1.38 $\pm$ 15 \% \\
\hline
$\Gamma_{\rm n}/\Gamma_{\rm p}$ & 0.46 & 0.23 $\pm$ 20 \% \\ 
\hline
$\Lambda \Lambda \to \Lambda n$ & 3.6 $\times$ 10$^{-2}$ &
1.2 $\times$ 10$^{-2} \pm$ 50  \% \\
\hline
$\Lambda \Lambda \to \Sigma^0 n$ & 1.0 $\times$ 10$^{-3}$ &
1.7 $\times$ 10$^{-3} \pm$ 30 \% \\ 
$\Lambda \Lambda \to \Sigma^- p$ & 2.0 $\times$ 10$^{-3}$ & 
3.4 $\times$ 10$^{-3} \pm$ 30 \% \\
\hline
$\Lambda \Lambda \to YN$ & 3.9 $\times$ 10$^{-2} $ 
& 1.7 $\times$ 10$^{-2}  \pm$ 30 \% \\ 
\hline
\end{tabular}
\label{tab:summarize}
\end{table}         

\end{document}